\documentclass[aps,pre,twocolumn,showpacs,superscriptaddress,groupedaddress]{revtex4}  
\usepackage{graphicx}  
\usepackage{float}
\usepackage{dcolumn}   
\usepackage{bm}        
\usepackage{amssymb}   
\usepackage{amsmath}
\usepackage{isomath}
\usepackage[usenames, dvipsnames]{color}
\usepackage{color,soul}

\begin{document}

\title{	Demonstration of dispersive rarefaction shocks in hollow elliptical cylinder chains}
\author{H. Kim}
\affiliation{Aeronautics and Astronautics, University of Washington, Seattle, WA, 98195-2400, USA}
\author{E. Kim}
\affiliation{Aeronautics and Astronautics, University of Washington, Seattle, WA, 98195-2400, USA}
\affiliation{Division of Mechanical System Engineering, Automotive Hi-Technology Research Center, Chonbuk National University,567 Baekje-daero, Deokjin-gu, Jeonju-si, Jeollabuk-do 54896, Republic of Korea}
\author{C. Chong}
\affiliation{Department of Mathematics, Bowdoin College, Brunswick, Maine 04011, USA}
\author{P. G. Kevrekidis}
\affiliation{Department of Mathematics and Statistics, University of Massachusetts, Amherst, MA, 01003, USA}
\author{J. Yang}
\affiliation{Aeronautics and Astronautics, University of Washington, Seattle, WA, 98195-2400, USA}

\date{\today}

\begin{abstract}
We report an experimental and numerical demonstration of dispersive rarefaction shocks (DRS) in a 3D-printed soft chain of hollow elliptical cylinders.
We find that, in contrast to conventional nonlinear waves, these DRS have their lower amplitude components travel faster, while the higher amplitude ones propagate slower. This results in the backward-tilted shape of the front
of the wave (the \textit{rarefaction} segment)
and the breakage of wave tails into a modulated
waveform (the \textit{dispersive shock} segment). 
Examining the DRS under various impact conditions, we find the counter-intuitive feature that the higher striker velocity causes the slower propagation of the DRS.
These unique features can be useful for mitigating impact controllably and efficiently without relying on material damping or plasticity effects.
\end{abstract}
\date{\today}
\maketitle

In recent decades, computational and experimental investigation of mechanical waves propagating in nonlinear lattices has been a subject of intense research. Primary efforts have been placed on exploring
solitary traveling waves~\cite{c7,c8} and discrete breathers~\cite{c10,c10a}; see also~\cite{GC, AVF}.
Arguably, less attention has been paid to the possibility of shock
wave formation, especially so at the experimental level within the
realm of granular crystals and mechanical metamaterials~\cite{c15, c14, McDonald}.
Herbold and Nesterenko investigated the formation of shock wave structures under the influence of viscous dissipation~\cite{c15}. Molinari \emph{et al.} \cite{c14} studied dispersive shock waves in uniform and
  periodic heterogeneous granular crystals, which feature oscillatory wave tails following the steady shock front. Shocks in disordered granular crystals were also studied in~\cite{Gomez}. In these studies, granular lattice elements interact with each other under the effective strain-hardening power law (i.e., compressive force $F$ and displacement $\delta$ have $F\sim\delta^p$ where the nonlinear exponent $p>1$)~\cite{c4}. 

If a discrete system can exhibit effective strain-softening behaviors ($p<1$), we can anticipate the emergence of distinctive features in comparison to the case of $p>1$. For instance, Herbold \textit{et al.} reported theoretical observation of rarefaction waves, which form tensile wavefronts despite the application of compressive impact \cite{Herbold}. More recently, Yasuda \textit{et al.} demonstrated numerically the formation of waves that combine a dispersive shock tail and a rarefaction front wave, so-called \textit{dispersive rarefaction shocks} (DRS), by using generalized
power-law contact models \cite{c2}. These studies, however, have been conducted without experimental verification,
though the experimental feasibility of such a setting has been discussed in tensegrity~\cite{tensegrity} and origami~\cite{origami} platforms.
If we can realize a physical system that supports the DRS, it would enable a two-fold efficient impact mitigation system for attenuating stress waves: one by transitioning the steep shock wavefront into a back-tilted form (within the \textit{rarefaction}) and the other by distributing energy over the space domain (within the \textit{dispersive shock}).

Soft materials have been emerging as a new playground for the formation of nonlinear waves \cite{c29, Raney}. The nonlinear behavior of soft materials depends not only on the mechanical properties of their constitutive materials, but also on their assembling architectures, e.g., geometrical configurations and buckling behavior \cite{c30, c32}. This offers us an enhanced degree of design freedom compared to conventional lattices, such as granular crystals whose tunability relies heavily on their local contact mechanics. The use of soft materials can be further beneficial for impact mitigation purposes, since their material damping efficiency can be enhanced on top of the intrinsic stress wave attenuation via wave dynamics effects, such as
  dispersion and rarefaction mentioned above.


In this Letter, we combine the above functionalities by deploying a soft-lattice system as a prototypical testbed for an experimental manifestation and corresponding numerical
  modeling of the DRS. Specifically, we fabricate a 3D-printed chain of hollow elliptical cylinders (HECs), and show that this nonlinear waveguide follows the strain-softening behavior with the nonlinear exponent $p<1$, stemming from its geometrical nonlinearity. Using this one-dimensional (1D) HEC chain, we demonstrate the emergence of the DRS under a striker impact condition. Furthermore, two of unique features of the DRS, i.e., the back-tilted wavefront due to the rarefaction and the
oscillatory wave tails due to the dispersive shock, are validated not only experimentally, but also computationally by using the finite element and the discrete element methods. Ultimately, we assess the effect of wave dispersion and rarefaction by the DRS in comparison to the material damping effect, confirming the efficacy of the HEC in stress wave attenuation.
\begin{figure}[t]
 \begin{center}
 \includegraphics[width=1\linewidth]{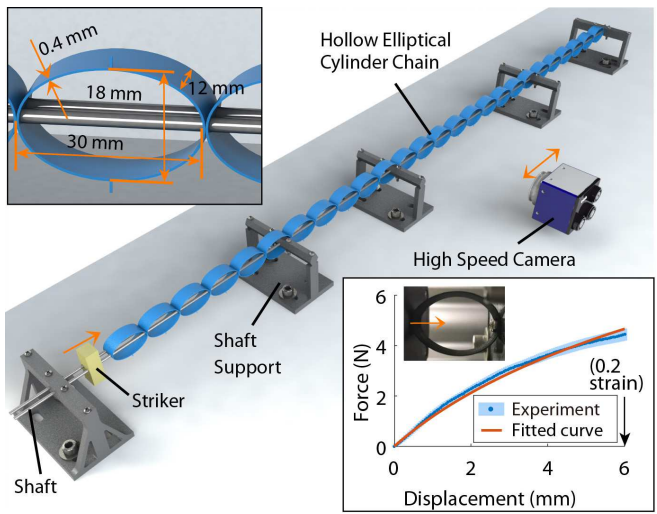}
 \end{center}
 \caption{A schematic diagram of the dynamic test setup. The top left inset shows the enlarged HEC unit cell and its dimensions. The bottom right inset is the force-displacement curve obtained from the quasi-static compression test plotted in the blue curve, along with the red solid curve that depicts the power-law fitting result. The digital image shows the compression test setup. 
}
 \label{fig1:exp_setup}
\end{figure}
The experimental setup is composed of a chain of HECs, a striker impact system, and measurement devices (Fig. \ref{fig1:exp_setup}). The chain consists of 26 HECs, which are 3D-printed (Ultimaker 3) with a poly-lactic acid (PLA) material and epoxy-bonded together at their interfaces. Each HEC has outer dimensions of 30 mm and 18 mm along the major and minor axes, respectively. The width and thickness are 12 mm and 0.4 mm (see the top inset of Fig. \ref{fig1:exp_setup}). The measured mass of each HEC is $m$ = 0.455 $\pm$ 0.006 g. Two linear stainless steel shafts (diameter: 2.38 mm) penetrate the side surfaces of the HECs to align them and to restrict their lateral motions. The two shafts are supported firmly by the 3D-printed jigs to minimize their vibrations. We note in passing that the HECs in this horizontal setup interact with each other following the power law with $p < 1$, which is confirmed by the quasi-static loading test (see the bottom inset of Fig. \ref{fig1:exp_setup} and Supplemental Material (SM)~\cite{c3} for details). 

To apply impact to the HEC system, we use a vibration shaker (LDS V406, B\&K) that launches a rectangular striker (PLA, mass: $m_s$ = 4.3 g) towards the first HEC in the chain at a controllable and consistent speed ($v_s=2.73\pm0.05$ m/s). The striker impact triggers the high speed camera (Phantom v1211) by means of a piezoelectric disc attached to the outer surface of the first HEC. The high speed camera is translated along the linear stage (BiSlide, Velmex) to capture the dynamic displacement profiles of each cylinder (i.e., $x_n$ for the $n$th particle) by using the digital image correlation technique~\cite{c3}. In each particle spot, we run the impact experiment five times for statistical treatment. 

\begin{figure}[t]
 \begin{center}
 \includegraphics[width=0.9\linewidth]{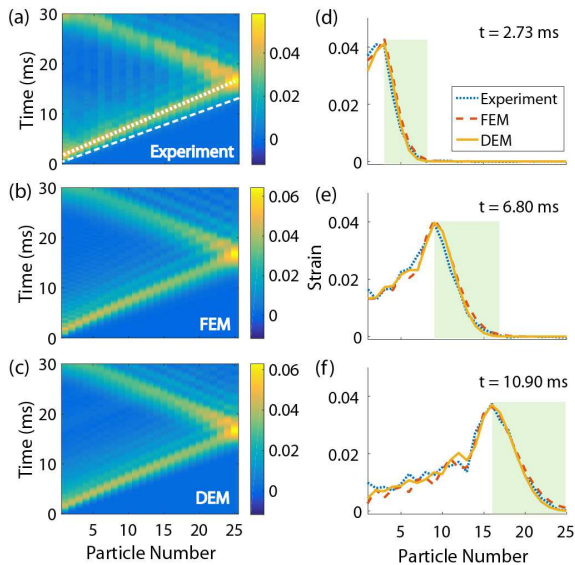}
 \end{center}
 \caption{Surface map of strains in time and space domains from: (a) Experiments, (b) Finite element method (FEM), and (c) Discrete element method (DEM). In (a), the white dotted line is the peak trace of the leading wave, and the white dashed line is the leading edge. Strain profiles are plotted at different time points, (d) $t$ = 2.73 ms, (e) 6.78 ms, and (f) 10.88 ms based on experiments (blue dotted curves), FEM (red dashed curves), and DEM (yellow solid curves). The green shaded area denotes the enlarging spatial width of the DRS wavefront.
}
 \label{fig2:exp_result}
\end{figure}

Based on the measured displacements, the strain between neighboring particles can be obtained as $u_n=(x_{n+1}-x_n)/a$, where $a$ is the major axis length of the cylinder ($a$ = 30 mm in this study). Figure \ref{fig2:exp_result}(a) shows the surface map of the measured strains in space and time domains based on the experimental data, see SM \cite{c3} for details. A unique feature to notice is that the shape of the leading pulse changes from the initial compactly-supported shape to a wider one (see the increasing gap between the front edge (dashed line) and the peak points (dotted line) in Fig. \ref{fig2:exp_result}(a), see also~\cite{linearDisp}). This implies that the wave component with the smaller amplitude (i.e., front edge) travels faster than the one with the larger amplitude (wave peak). This results in the deformation of the waveform, such that it gradually leans backward and shifts the wave peak location to the rear. This is more evident from Figs. \ref{fig2:exp_result}(d-f), where the shaded areas show the evolution of the wavefront shape in the space domain over time. Corroborated by the numerical observations discussed below, these experimental results definitively showcase the formation of the rarefaction wavefront.
 
\begin{figure}[t]
 \begin{center}
 \includegraphics[width=1.0\linewidth]{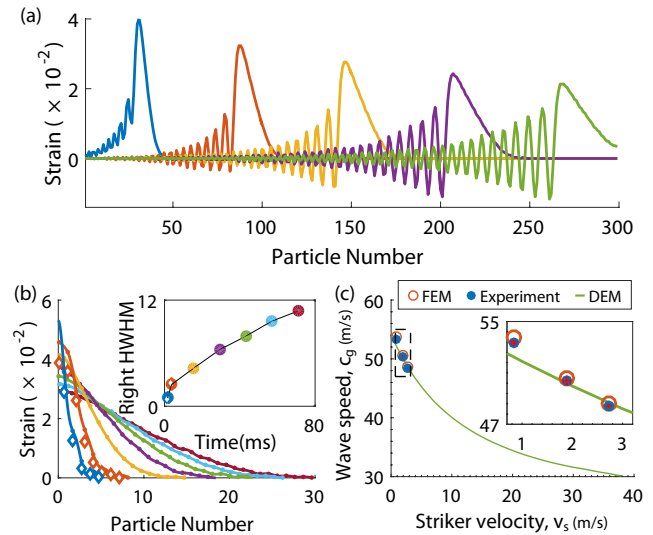}
 \end{center}
 \caption{(a) FEA striker impact simulation result with the striker velocity $v_s=2.73$ m/s for a long ($N=300$) chain, no damping included. The strain is calculated at time $t$ = 20, 52, 84, 116, 148 ms from left to right. (b) The evolution of the leading pulse's waveform (right halves shifted to the origin) over time ($t$ = 1.7, 3.9, 15, 29, 43, 57, and 71 ms from compact- to broad-supported shapes). The inset shows the right half-width-at-half-maximum (HWHM) over time. The diamond markers denote experiment results. (c) Group wave speed ($c_g$) of DRS as a function of the striker velocity ($v_s$) for the HEC chain with $N=26$. The green curve shows DEM simulations for the parametric study of $c_g$ with respect to $v_s$, while the experimental and FEA results are shown in solid blue and hollow red circles, respectively, for three different striker velocities ($v_s \approx$ 0.9, 1.8, and 2.7 m/s). The inset shows an enlarged view around the experiment data points. The negligible red bars on top of the experiment data indicate standard deviations in striker velocity (in the $x$-direction) and the wave speed (in the $y$-direction). 
}
 \label{fig3}
\end{figure}

Another feature to take note of from Fig. \ref{fig2:exp_result}(a) is that the wave shows oscillating wave trails whose peak amplitudes feature a monotonic decrease [see also the trailing part of the wave in Fig. \ref{fig2:exp_result}(f)]. This oscillatory pattern combining the amplitude-dependent wave speed with the manifestation of dispersive phenomena is the principal characteristic of dispersive shock waves \cite{c6}.
In addition to the oscillatory wave tail, the aforementioned back-tilted wavefront enables us to confirm the experimental verification of the DRS in the HEC system (further details including frequency analysis are provided in SM~\cite{c3}).
We also conduct numerical simulations of the DRS by using a  finite element method (FEM) [Fig. \ref{fig2:exp_result}(b), see SM~\cite{c3} for details]. The formation of the DRS is also evident in the FEM results, and the DRS profiles based on the FEM are in agreement with the experimental results [compare Figs. \ref{fig2:exp_result}(a) and (b), and also see the spatial waveforms in Figs. \ref{fig2:exp_result}(d-f)]. The advantage of the FEM is that we can extend the chain length at will, so that we can observe the evolution of the DRS over a larger space domain, which, in turn, enables a more pronounced manifestation of the relevant phenomenology. Figure~\ref{fig3}(a) shows the FEM simulation result of the spatial wave profiles of the DRS using an HEC chain with $N=300$. First, we can clearly observe that the wave tail develops into a modulated waveform as the wave propagates through the HEC chain. This is strongly reminiscent of the multi-scale manifestation of dispersive shock waves in different contexts~\cite{c5}. In particular, the fast-traveling oscillatory waves which are harmonic when viewed in a local scale, bear an envelope of a slowly decaying modulation in a larger scale.
Analyzing the frequency components in the wave tails, we find that they follow the local resonance of the HEC derived from its \textit{nonlocal} geometry (see details in SM~\cite{c3}). This is different from the conventional discrete lattice systems, whose dynamics is highly dependent on their \textit{local} contact mechanics, rather than their soft constitutive mechanics. 
 
To complement the analysis of the wave, Figure \ref{fig3}(b) shows the leading pulse profiles of the DRS, collected at different temporal moments but all aligned with respect to the origin in the space domain. It is evident that the wavefront width expands while its peak is attenuated. The evolution of the wave width is quantified in the inset in terms of the half-width-at-half-maximum (HWHM), which shows a monotonic increase. The experimental data points in hollow markers corroborate the FEM results for the short-chain region. 


While the FEM provides us with an accurate computational visualization of the experimental phenomenology, it would be beneficial to derive a simple yet effective model of the HEC chain. With a proper model capturing the principal features of the dynamics discussed herein, we can enhance our understanding of the forming mechanism of the DRS. To this end, we approximate the continuum HEC system via a 1D monomer chain of lumped masses based on the discrete element method (DEM). In this discrete system, the neighboring HEC particles are assumed to interact with each other by the following power-law:
\begin{equation}\label{eqn:Hertz}
F=A({\Delta} x+\delta_0)^p-f_0,
\end{equation} 
where $F$ is the contact force, $\Delta x$ is the relative
displacement between neighboring HEC centers, $\delta_0$ is an effective pre-compression term, $p$ is the nonlinear exponential between the HECs, and $f_0$ is a force constant to incur no interactions under zero particle displacement (i.e., $f_0 = A \delta_0 ^ p$). The validity of this power law in our HEC system is demonstrated by the curve fitting result shown in the bottom inset of Fig.~\ref{fig1:exp_setup} (further details for deriving the coefficients of Eq.~\eqref{eqn:Hertz} are in SM~\cite{c3}).


For the $n$th particle in the HEC chain, the equation of motion  can be written as
\begin{equation}\label{eom}
m\ddot{x}_n = A(\delta_0+x_{n-1}-x_n)^p-A(\delta_0+x_n-x_{n+1})^p-c_d\dot{x}_n,
\end{equation}
where $n=2,3,\cdots,N-1$ ($N=26$), the overdot denotes a time-derivative, and $c_d$ is the damping coefficient obtained empirically by curve-fitting with the experimental data. To account for the boundary conditions, the equations of motions for the first ($n$ = 1) and last ($n$ = N) particles need to be modified (details in SM~\cite{c3}). We solve these differential equations using the fourth-order Runge-Kutta routine to analyze the dynamic response of the discretized HEC chain. Note that for the accurate comparison with the experimental results, we feed into the solver the first particle's displacement profile (i.e., $x_1$) obtained from the experiment. As a result, the strain surface map based on the DEM is plotted in Fig. \ref{fig2:exp_result}(c). The DEM result is in good agreement with the experimental one [Fig. \ref{fig2:exp_result}(a)]. The spatial profiles of the propagating DRS also corroborate those from the experiments and the FEM [Figs. \ref{fig2:exp_result}(d-f)].

By leveraging the fast and efficient computation of the DEM, we move on to the next question; as the striker velocity ($v_s$) is varied, how  will the resulting DRS be affected as characterized by  its group velocity ($c_g$)?
Conventional nonlinear waves, including shocks, tend to generate faster traveling waves as we impose higher external excitations. For example, Nesterenko \cite{c7} derived the relationship $c_g \sim v_p^{1/5}$ in granular crystals, where $v_p$ is the particle velocity that is directly related to $v_s$. 
Figure \ref{fig3}(c) shows the DEM calculation of $c_g$ (green curve) as a function of $v_s$, where $c_g$ is obtained by tracing the peak points in the strain map and calculating their averaged slope [e.g., see the dotted-line slope of Fig.~\ref{fig2:exp_result}(a)]. Note that given the short chain ($N$ = 26), the variation of $c_g$ along the chain is less than 1\%. In Fig.~\ref{fig3}(c), it is striking that the leading pulse of the DRS
propagates {\it slower} for \textit{higher} external excitations in terms of the striker velocity applied. 

To experimentally verify this important by-product of the strain-softening nature of the HEC chain, we conduct impact tests with various striker velocities: $v_s = 1.89\pm0.07$ m/s and $v_s = 0.83\pm0.04$ m/s, which are roughly 2/3 and 1/3 of the previous striker velocity. Their results are plotted in Fig. \ref{fig3}(c) in solid dots (see also the inset, where error bars based on five tests are almost invisible due to consistency). The numerical results based on the FEM are also marked in red circular dots. We confirm that the experimental and computational results corroborate the negative correlation between $v_s$ and $c_g$ predicted by the DEM. We also note that at $v_s\approx$ 0, we have $c_g \approx 54$ m/s, which corresponds to the linear wave (i.e., sound wave) speed in the HEC chain. 
Becasue of this asymptotic nature of $c_g$, we find that the relationship between the group and phase speeds does not follow the power law of $c_g \sim v_p^{1/5}$ unlike the typical granular chains (see SM \cite{c3} for details).

\begin{figure}[t]
 \begin{center}
  \includegraphics[width=0.9\linewidth]{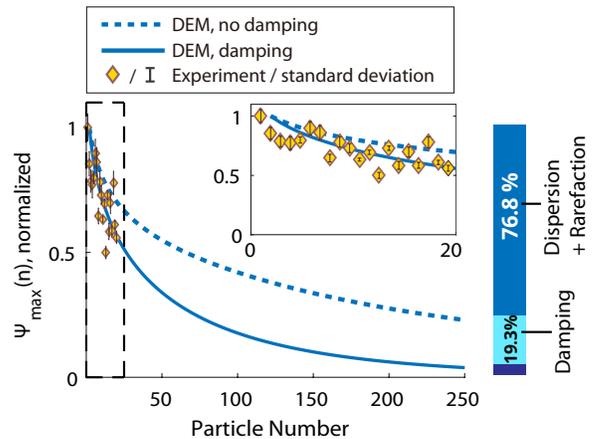}
 \end{center}
 \caption{The peak potential energy for each cylinder, normalized by the input potential energy, is plotted as a function of the inter-particle location. Solid and dotted curves represent the DEM results with and without damping, respectively. Experiment results are plotted as yellow diamonds in the beginning of the chain up to $N=20$, which can be more clearly seen in the inset (enlarged plot of the dashed area). The bar graph on the right shows the contribution of the damping (cyan) and the combined effect of dispersion and rarefaction (blue) to the overall potential energy reduction (in case of $c_d=0.003$ and $c_s=0.2$).}
 \label{fig4}
\end{figure}

To assess the efficacy of the HEC chain as an impact mitigation system, we calculate the evolution of the maximum potential energy experienced by each inter-particle location, as the wave propagates along the chain. The potential energy $\Psi(n)$ stored in the $n$th inter-particle spot can be simply found by integrating Eq.~\eqref{eqn:Hertz} as
\begin{multline*}
\Psi(n) =  \dfrac{A}{p+1}\left[(\delta_0+x_n-x_{n+1})^{p+1}-\delta_0^{p+1}\right]
\\ -f_0(x_n-x_{n+1}).
\end{multline*}
\noindent We calculate the potential energy over time and find a peak value, $\Psi_\text{max}(n)$, in each inter-particle location. This potential energy value after normalization is shown in Fig. \ref{fig4}. The solid curve denotes the DEM results based on the curve-fitting with the experimental data (see diamond dots with error bars in the inset). In this process, the degree of the material damping -- in terms of the chain damping coefficient $c_d$ (in Eq.~\eqref{eom}) and the striker damping coefficient $c_s$ (see SM~\cite{c3}) -- is optimized, such that the DEM best fits the experimental trend. We observe the decay of the peak potential energy over the spatial regime, which manifests a highly efficient mechanism of stress wave attenuation in the HEC system. 

It is now natural to inquire about the portion of this attenuation contributed by the combined dispersion and rarefaction mechanism in the DRS, compared to the material damping effect. This question can be answered by assessing the effect of the damping on the overall wave attenuation. For this, we run the DEM simulation with zero damping coefficients. The results are shown in the dotted curve in Fig. \ref{fig4}, which also shows a rapid drop of $\Psi(n)$ over the space. Comparing the two DEM cases (i.e., solid and dashed curves), the energy reduction from the non-damped to the damped DEM results is 19.3\% over the span of 250-particle chain. However, the potential energy drop even for the non-damped case is around 76.8\% at the end of the chain, compared to the initial energy level. This implies that the wave attenuation solely due to the combined dispersion and rarefaction is more than three times larger than that due to the damping in the given system. 
Though the relative portions can change depending on the system configurations, size, and boundary conditions, this trend overall supports that the formation of the DRS can be an efficient way of mitigating stress waves without resorting to material damping or plasticity effects.



In summary, we observed the dispersive rarefaction shock (DRS) dynamics in the soft chain of 3D-printed hollow elliptical cylinders (HECs). We experimentally and numerically validated the two principal features of the DRS, the back-tilted wavefront in the form of a rarefaction and the oscillatory wave tail in the form of a dispersive shock.
Moreover, we demonstrated that the HEC system supports a slower propagation of DRS given a higher striker impact condition, as a result of the strain-softening nature of this nonlinear dynamical lattice.
The proposed HEC system can be potentially applied to the impact mitigation system design in various scientific and engineering applications: our results clearly manifested its efficiency in spreading the originally  stored potential energy during propagation.
Further research can be pursued by modifying the discrete element model (DEM) by adding more degrees of freedom to capture the higher modes of wave propagation. Indeed, as discussed in SM~\cite{c3}, while the phenomenology presented here hinges on the lowest band of its dynamics, the HEC lattice bears intriguing characteristics associated with multiple bands and gaps that are certainly worthwhile of additional exploration.
Future studies also include investigating the role of defects (e.g., breather formation) in strain-softening systems. The systematic development of the HEC as a prototypical strain-softening element may also pave the way for exploring heterogeneous chains involving the alternation of softening and hardening nonlinearities, which may, in turn, manifest unprecedented nonlinear phenomena.

We would like to acknowledge the financial support from the NSF (CAREER-1553202 and DMS-1615037). We give special thanks to Dennis Wise and Carl Knowlen at the University of Washington, Waverly Harden at Bowdoin College, and Pai Wang at Harvard University for technical assistance.  E. K. acknowledges the support from the National Research Foundation of Korea (NRF) grant funded by the Korea government (MSIP, No. 2017R1C1B5018136). P. G. K. and J. Y. gratefully appreciate support from the AFOSR via the DDDAS program (FA9550-17-1-0114).



\begin{thebibliography}{2}
\bibitem{linearDisp} The change of the wave width from compact- to wide-support in the HEC system stems from the nonlinearity, which should be distinguished from the conventional wave dispersion in linear elastic media. Details on the backward-tilted shock formation are in [13].
\end{thebibliography}
\end{document}